\documentclass[aps,twocolumn,pra,superscriptaddress,showpacs,tightenlines]{revtex4}
\usepackage{amssymb}
\usepackage{amsmath}
\usepackage{graphicx}
\usepackage{amsfonts}

\begin{document}

\title{Coherent output of photons from coupled superconducting transmission
line resonators controlled by charge qubits}
\author{Lan Zhou}
\affiliation{Department of Physics, Hunan Normal University, Changsha 410081, China}
\affiliation{Institute of Theoretical Physics, Chinese Academy of Sciences, Beijing,
100080, China}
\author{Y.B. Gao}
\affiliation{College of Applied Sciences, Beijing University of Technology, Beijing,
100022, China}
\author{Z. Song}
\affiliation{Department of Physics, Nankai University, Tianjin 300071, China}
\author{C.P. Sun}
\affiliation{Institute of Theoretical Physics, Chinese Academy of Sciences, Beijing,
100080, China}
\affiliation{Department of Physics, Nankai University, Tianjin 300071, China}
\email{suncp@itp.ac.cn}
\homepage{http://www.itp.ac.cn/~suncp}

\begin{abstract}
We study the coherent control of microwave photons propagating in a
superconducting waveguide consisting of coupled transmission line
resonators, each of which is connected to a tunable charge qubit. While
these coupled line resonators form an artificial photonic crystal with an
engineered photonic band structure, the charge qubits collectively behave as
spin waves in the low excitation limit, which modify the band-gap structure
to slow and stop the microwave propagation. The conceptual exploration here
suggests an electromagnetically controlled quantum device based on the
on-chip circuit QED for the coherent manipulation of photons, such as the
dynamic creation of laser-like output from the waveguide by pumping the
artificial atoms for population inversion.
\end{abstract}

\maketitle

\section{\label{sec:one}Introduction}

Recent experiments with on-chip all optical setups~\cite{Fan1,Fan2,Fan3,Fan4}
have displayed slow light phenomenon similar to that due to the
electromagnetically induced transparency (EIT)~\cite{EIT97, EIT01}. Here, a
physical mechanism was is presented using a model with a coupled resonator
optical waveguide, which behaves as a photonic crystal with a band-gap
spectrum. The coupling of each resonator to some external cavities can shift
the resonant spectral line and compress the bandwidth, and thus stop or
store the propagating light pulses~\cite{Fan1,Fan2,Fan3}.

\begin{figure}[tbp]
\includegraphics[bb=80 250 520 610, width=7 cm, clip]{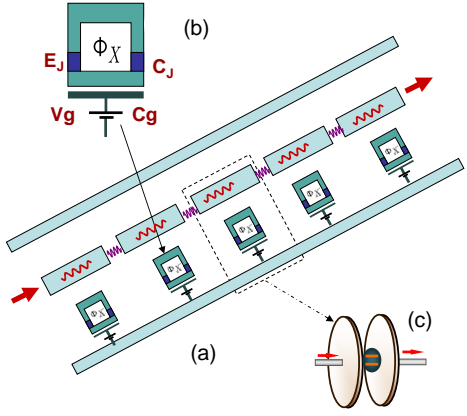}
\caption{(Color online) Configuration of the setup for controlling light
propagation in a coupled-line-resonator waveguide (a) by coupling to charge
qubits (b). The coupled (line resonators) -- (charge qubits) system can
behave similarly to the cavity QED for a single atom interacting with a
single-mode cavity (c).}
\label{fig:1}
\end{figure}

Motivated by this progress, both in experimental and theoretical aspects, we
propose and study a hybrid structure with a cavity waveguide interacting
with two-level artificial atoms. Here, the novel control mechanism for
coherent transmission of microwave photons in the waveguide is to utilize
the collective excitations of the atoms, which can be described as
quasi-spin waves~\cite{sunprl91} in the low excitation limit. As illustrated
in Fig.~\ref{fig:1}, we suggest a co-planar on-chip setup based on the
superconducting circuit QED~\cite%
{YouPT58,YouPRB68,YouJPRB68,NTTPRL96,mn431,yale1,yale2,liuPRA71,liuEL67}:
each cavity in our proposed setup has been experimentally implemented as a
superconducting transmission line resonator; the spatially distributed
artificial atoms are the biased Cooper pair boxes (charge qubits), which
play the same role as the external cavity for controlling light in the
waveguide of the on-chip all optical experiment~\cite{Fan1,Fan2,Fan3,Fan4}.

Due to its engineered photonic band structure, such a cavity array, coupled
to quasi-spin wave excitations, can result in much richer quantum coherent
phenomena. We show that the quasi-spin waves of charge qubits can
controllably affect the engineered photonic band structure so that the
modified dispersion relation results in a slow (and even zero) group
velocity of photons propagating along the waveguide of coupled line
resonators. The spin-wave excitation can also drive a coherent state of
photons in the coupled resonator waveguide. Some pumping methods can make a
population inversion to produce a laser-like output above the threshold.

This paper is organized as follows. In Sec.~\ref{sec:two}, we present our
hybrid system: an array of coupled-line-resonators based on a
superconducting circuit. Each resonator is coupled to a biased Cooper pair
box (CPB). The array of coupled-line-resonator exhibits a band structure. In
Sec.~\ref{sec:three}, in the large-$N$ and low-excitation limits, the hybrid
system is modeled as two coupled boson models, with one for the photonic
band and the other for the spin wave of $N$ CPBs. Here, the slow light
phenomenon is found by controlling the detuning and the coupling strength.
The dynamic creation of the laser-like output from the array of
coupled-line-resonator is proposed in Sec.~\ref{sec:four} and~\ref{sec:five}%
. This is because two kind of bosons couple linearly with each other and the
population inversion of the artificial atoms is pumped. Sec.~\ref{sec:six}
presents our conclusions.

\section{\label{sec:two}The coupled cavity QED based on superconducting
circuit}

Now we consider a superconducting quantum circuit including $N$ CPBs and an
array of coupled-line-resonator. Recently, most experiments have
demonstrated the possibility to fabricate a superconducting qubit array~\cite%
{nec2q1,nec2q2, 4q}. As shown in Fig.~\ref{fig:1}, the array of
coupled-line-resonators is constructed by cutting the superconductor
transmission line into $N$ segments. And the $N$ CPBs are made of $N$ dc
SQUIDs (superconducting quantum interference device) which consists of two
tunnel junctions.

With a proper biased voltage $V_{g}$, each CPB behaves as a two-level system
(charge qubit)~\cite{nec99}. Typically, the model Hamiltonian for a charge
qubit can be written as in Ref.~\cite{YouPT58,GWe0508}
\begin{equation}
H=\frac{1}{2}B_{z}\sigma _{z}-\frac{1}{2}B_{x}\sigma _{x}
\end{equation}%
with the level spacing of the charge qubit
\begin{equation}
B_{z}=4E_{C}\left( 2n_{g}-1\right)
\end{equation}%
and effective Josephson energy
\begin{equation}
B_{x}=2E_{J}\cos \left( \pi \frac{\Phi _{x}}{\Phi _{0}}\right) .  \label{?}
\end{equation}%
The quasi-spin operators $\sigma _{z}$ and $\sigma _{x}$ are defined in the
charge qubit basis ($\left\vert 0\right\rangle $ and $\left\vert
1\right\rangle $), where $0$ and $1$ represent excess Cooper pairs on the
superconducting island, respectively. Here, $E_{C}$ and $E_{J}$ represent
the charge energy and the Josephson energy, respectively. Through $%
n_{g}=C_{g}V_{g}/2e$, we can tune the degenerate point of the charge qubit
by controlling the gate voltage $V_{g}$ applied on the gate capacitance $%
C_{g}$. In addition, $\Phi _{x}$ represents the magnetic flux through the
SUID loop induced by the externally applied magnetic field and $\Phi
_{0}=h/2e$ is the flux quantum.

Based on the above results, we can write the Hamiltonian including $N$ CPBs
as
\begin{equation}
H_{A}=\frac{\omega _{A}}{2}\sum_{j}\left( |e_{j}\rangle \langle
e_{j}|-|g_{j}\rangle \langle g_{j}|\right)
\end{equation}%
where the eigen-frequency $\omega _{A}=\sqrt{B_{z}^{2}+B_{x}^{2}}$. Based on
recent experiments, the charge energy can be taken as $E_{C}=29.5$ GHz,
while the Josephson energy $E_{J}=8$ GHz. For simplicity, all qubits are
assumed to be identical and biased off the degenerate point. The energy
eigenstates
\begin{equation}
\left\vert e_{j}\right\rangle =\cos \left( \frac{\theta }{2}\right)
|0_{j}\rangle -\sin \left( \frac{\theta }{2}\right) |1_{j}\rangle
\label{ctl2-02}
\end{equation}%
and
\begin{equation}
\left\vert g_{j}\right\rangle =\sin \left( \frac{\theta }{2}\right)
|0_{j}\rangle +\cos \left( \frac{\theta }{2}\right) |1_{j}\rangle \text{,}
\label{ctl2-03}
\end{equation}%
are the superpositions of charge eigenstates $\left\vert 0_{j}\right\rangle $
and $\left\vert 1_{j}\right\rangle $, and $\theta =\arctan \left(
B_{x}/B_{z}\right) $.

In our setup, each qubit is placed at the position of the anti-node of the
standing wave field in each transmission line resonator. The London equation
provides the vanishing boundary conditions for the quantized electromagnetic
field at the two ends of each line resonator. Thus, the quantized magnetic
field vanishes at those anti-nodes and the qubits are only coupled to the
electric component. Here, the gate voltage applied to the capacitance of $j$%
th Cooper pair box is given by $V_{x}^{j}=V_{g}+V_{q}^{j}$, where $V_{q}^{j}$
stands for the quantized part of the voltage. Let only the $n=2$ mode be
activated. Then the gate voltage is quantized as \cite{yale1,yale2}
\begin{equation}
V_{q}^{j}=(\hat{a}_{j}+\hat{a}_{j}^{\dag })\sqrt{\frac{\omega }{LC}}
\end{equation}%
at its maximum while the magnetic field vanishes. Here, we only assumed a
single mode of the quantized electric field of frequency $\omega $ with
creation (annihilation) operator $\hat{a}_{j}^{\dag }$ ($\hat{a}_{j}$). $L$
is the length of the line resonators and $C$ is the capacitance per unit
length.

There exists the coupling $J$ between two neighbor line resonators through
the dielectric material. Then the Hamiltonian of the coupled line resonators
waveguide is
\begin{equation}
H_{C}=\omega \sum_{j}\hat{a}_{j}^{\dag }\hat{a}_{j}+J\sum_{j}\left( \hat{a}%
_{j}^{\dag }\hat{a}_{j+1}+\hat{a}_{j+1}^{\dag }\hat{a}_{j}\right) \text{,}
\end{equation}%
where $J$ depends on the coupling mechanism, e.g., the magnetic penetration
depth.

Following recent experiments~\cite{yale1}, we can take the frequency of the
line resonators $\omega =10.0$ GHz, the length of line resonators $L=1.0$ cm
and the capacitance per unit length $C=0.13$ fF$/\mu $m. We now make the
rotating wave approximation to write down the interaction Hamiltonian~\cite%
{yale1,yale2}
\begin{equation}
H_{I}=g\sum\limits_{j=1}^{N}\left( \hat{a}_{j}|e_{j}\rangle \langle g_{j}|+%
\text{H.c.}\right) ,
\end{equation}%
between the qubits and the fields, where
\begin{equation}
g=e\sin \theta \frac{C_{g}}{C_{\Sigma }}\sqrt{\frac{\omega }{LC}}
\end{equation}%
is the coupling strength between the resonator and the artificial two-level
atom, and $C_{\Sigma }$ is the sum of the gate capacitance and the
capacitance of the tunnel junction.

In practical experiments, the coupling constants $g$ and $J$ should depend
on the position of the qubit. For simplicity, in this paper we take a
uniform $g$ and $J$. Theoretically, the fluctuations of the coupling
constants are assumed to be innocuous and do not change the results of this
paper, qualitatively. The above model has been used to demonstrate the
photon blocked effect~\cite{Bqp06159, Pqp06097,Dtree}, which leads to the
Mott insulating effect for the polaritons formed by dressing atoms with a
cavity field.

Although our proposed configuration setup is an extension of the single
cavity proposed in~\cite{YouPRB68,YouJPRB68,yale1,yale2} to N coupled
cavities, there are significant difference between them. In experiments~\cite%
{mn431,yale1}, to demonstrate the typical cavity QED character of a
superconducting quantum circuit for quantum computing, only a single cavity
is used rather than a coupled cavity array. A single cavity possesses
discrete photon modes with large frequency spacings, but an array of
coupled-line-resonator possesses a band-gap spectrum and can transmit a wave
packet of light. A single cavity coupled to many charge qubits in a similar
superconducting circuit has been modeled~\cite{Wang} to probe the dynamic
behavior of quantum phase transition \cite{Quan}. Obviously, different from
the setup in~\cite{yale1,yale2}, our setup uses many coupled transmission
line resonators to realize an electromagnetically controlled quantum device
for coherent control of photon transmission.

\section{\label{sec:three}Two-mode boson model for the spin-wave dressed
photonic band}

Now we consider the low-excitation limit that a few charge qubits are
populated in their excited state. The crucial issue here is to use the
collective operators~\cite{sunprl91}
\begin{equation}
\hat{B}_{k}^{\dag }=N^{-1/2}\sum_{j=1}^{N}\exp (ik\ell j)|e_{j}\rangle
\langle g_{j}|  \label{ctl3-01}
\end{equation}%
and its conjugate $\hat{B}_{k}=(\hat{B}_{k}^{\dag })^{\dag }$ to describe
the spin wave excitation of the charge qubit array, where $k=2\pi n/\ell N$
with $n=0,1,...,N-1$. In the large $N$ limit with low excitations, i.e., $%
\langle \sum_{k}\hat{B}_{k}^{\dag }\hat{B}_{k}\rangle \ll N$, these
collective excitations behave as bosons, since the usual bosonic commutation
relation $\left[ \hat{B}_{k},\hat{B}_{k}^{\dag }\right] =\delta _{kk^{\prime
}}$ can be approached when $N\rightarrow \infty $.

The Fourier transformation $\hat{a}_{k}=\sum_{j}\exp (ik\ell j)\hat{a}_{j}/%
\sqrt{N}$ diagonalizes the coupled resonator Hamiltonian as
\begin{equation*}
H_{C}=\sum_{k}\Omega _{k}\hat{a}_{k}^{\dag }\hat{a}_{k}
\end{equation*}
to give a dispersion relation with band structure
\begin{equation}
\Omega _{k}=\omega +2J\cos (k\ell ),  \label{ctl3-02}
\end{equation}%
where $\ell $ is the site distance. Then the normal modes of\ the hybrid
system with Hamiltonian $H=\sum_{k}H_{k}$\ are characterized by $\hat{a}_{k}$
and $\hat{b}_{k}=\lim_{N\rightarrow \infty }\hat{B}_{k}$. The evolution of
each mode is governed by the following Hamiltonian
\begin{align}
H_{k}& =\Omega _{k}\hat{a}_{k}^{\dag }\hat{a}_{k}+g\left( \hat{a}_{k}\hat{b}%
_{k}^{\dag }+\text{H.c.}\right) +\omega _{A}\hat{b}_{k}^{\dag }\hat{b}_{k}
\label{ctl3-03} \\
& =\Omega _{Dk}\hat{N}_{k}+\epsilon _{k}(\hat{P}_{k}^{\dag }\hat{P}_{k}-\hat{%
Q}_{k}^{\dag }\hat{Q}_{k}).  \notag
\end{align}%
Here, we have introduced the polariton operators~\cite{sunprl91}
\begin{align}
\hat{P}_{k}& =\cos \theta _{k}\hat{a}_{k}+\sin \theta _{k}\hat{b}_{k};
\label{ctl3-04} \\
\hat{Q}_{k}& =\sin \theta _{k}\hat{a}_{k}-\cos \theta _{k}\hat{b}_{k}.
\end{align}%
which are linear combinations of the quantized electromagnetic field
operators and atomic collective excitation operators of quasi-spin waves.
The mixing angle $\theta$ is determined by
\begin{equation}
\tan \theta _{k}=2g/(\Omega _{k}-\omega _{A}).
\end{equation}
The total excitation number for the $kth$ mode is given by
\begin{equation}
\hat{N}_{k}=\hat{P}_{k}^{\dag }\hat{P}_{k}+\hat{Q}_{k}^{\dag }\hat{Q}_{k}=%
\hat{a}_{k}^{\dag }\hat{a}_{k}+\hat{b}_{k}^{\dag }\hat{b}_{k}.
\label{ctl3-05}
\end{equation}%
The dispersion relation for the photonic band is obtained as
\begin{equation}
\epsilon _{\pm k}=\Omega _{Dk}\pm \varepsilon _{k},  \label{ctl3-06}
\end{equation}%
where
\begin{align}
\Omega _{Dk}& =\frac{1}{2}(\Omega _{k}+\omega _{A}),  \label{ctl3-07} \\
\varepsilon_{k}& =\frac{1}{2}\sqrt{(\Omega _{k}-\omega _{A})^{2}+4g^{2}}.
\notag
\end{align}%
Since $\hat{N}_{k}$ commutes with $H_{k}$, the number of excitations $\hat{N}%
_{k}$ is conserved, while the number of different type excitations $\hat{a}%
_{k}^{\dag }\hat{a}_{k}$ and $\hat{b}_{k}^{\dag }\hat{b}_{k}$ are mutually
convertible by adjusting the coupling strength $g$ and the detuning $\delta
=\omega -\omega _{A}$ between the artificial atom and the resonator. Due to
the coupling between artificial atoms and the resonators, the origin band
structure for the array of coupled-line-resonators is splitted into two in
the one excitation subspace. Obviously there exists a gap between two bands
for nonvanishing $g$.

Notice that the band structure (\ref{ctl3-07}) is only available in the low
excitation limit $\langle \sum_{k}\hat{b}_{k}^{\dag }\hat{b}_{k}\rangle \ll
N $. In the following, we focus on the single excitation subspace. For $%
\delta >0$, the corresponding dressed spectrum $\epsilon _{\pm k}=\Omega
_{Dk}\pm \epsilon _{k}$ shows a bandwidth narrowing effect on the coupled
line resonators due to its couplings to the charge qubits. This is because
the bandwidth $W_{-}=|\epsilon _{-k=0}-\epsilon _{-k=\pi }|$ of the low-band
changes from $2J$ to $2J-\Delta $, where
\begin{equation}
\Delta =A(J)-A(-J)
\end{equation}%
and
\begin{equation}
A(J)=\sqrt{(\frac{\delta }{2}+J)^{2}+g^{2}}.  \label{ctl3-08}
\end{equation}%
Meanwhile, the bandwidth $W_{+}=|\epsilon _{+k=0}-\epsilon _{+k=\pi }|$ of
the upper band changes from $2J$ to $2J+\Delta $. Obviously when $\delta =0$%
, the bandwidths $W_{-}$ and $W_{+}$ of the two bands are equal to $2J$;
when $\delta >0$ $(\delta <0)$, $W_{-}$ $(W_{+})$ is narrower than $2J$
while $W_{+}$ $(W_{-})$ is wider than $2J$. Then the wave packet of light in
the lower band can be adjusted by $\delta $. The couplings also shift the
central spectral line from $\omega $ and $\omega _{A}$ to $\epsilon _{\pm
k=\pi /2}$ respectively. This just recovers the same result obtained in Ref.~%
\cite{Fan1,Fan2,Fan3,Fan4}.

\begin{figure}[tbp]
\includegraphics[width=8.5 cm]{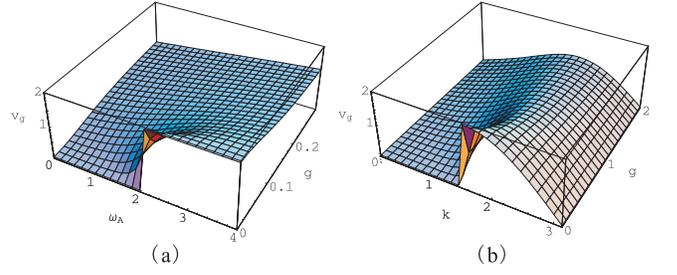}
\caption{(Color online) (a) the group velocity $v_{g}$ of the lowest
band as a function of $\protect\omega _{A}$ and coupling strength
$g$ at $k=\pi /(2\ell )$ for $\eta =\protect\gamma $, and
$\protect\omega =J$. It can be seen that when $\delta \gg 2g$, the
group velocity reaches its minimum $v_{g}(-\pi /(2\ell ))\approx 0$;
when $\delta\ll -2g$, the group velocity reaches its maximum
$v_{g}(-\protect\pi /(2\ell ))\approx 2J\ell $. (b) the group
velocity of the lowest band as a function of $k$ and coupling
strength $g$ in the case of resonance. Here, $g$ and $\omega _{A}$
are in units of $J$, and $k$ is in units of the lattice constant
$\ell$. } \label{fig:2}
\end{figure}

Then we can obtain the group velocities $v_{\pm }(k)=d\epsilon _{\pm k}/dk$
for the lower and the upper bands as
\begin{equation}
v_{\pm }=\text{Re}\left\{ J\ell \sin (k\ell )\left[ 1\pm \frac{\Lambda
+2J\cos (k\ell )}{\sqrt{[\Lambda +2J\cos (k\ell )]^{2}+4g^{2}}}\right]
\right\} ,  \label{ctl3-09}
\end{equation}%
where $\Lambda =\delta +i\kappa $. Here, $\kappa =\eta -\gamma $ is the
difference between the damping constant $\eta $ of each cavity and the decay
rate $\gamma $ of the qubit, where $\eta $ is caused by the cavity loss and $%
\gamma =T_{1}^{-1}+T_{2}^{-1}$: in experiments, the energy relaxation time $%
T_{1}$ of a superconducting qubit is 0.1 to a few microseconds, and the
dephasing time $T_{2}$ between the ground and the excited state is a few
dozens of nanoseconds. It can be seen that the group velocity becomes
independent of loss when $\eta =\gamma $. At the band center $k=\pi /(2\ell
) $, the group velocity for the lower band becomes
\begin{equation}
v_{-}\left( -\frac{\pi }{2\ell }\right) =\text{Re}\left\{ J\ell \left( 1-%
\frac{\Lambda }{\sqrt{\Lambda ^{2}+4g^{2}}}\right) \right\} \text{.}
\label{ctl3-10}
\end{equation}%
In the large detuning case, i.e., $\delta \gg 2|g|$, the group velocity
reaches its minimum
\begin{equation}
v_{-}(-\pi /(2\ell ))=\text{Re}\left[ J\ell g^{2}/\Lambda ^{2}\right]
\approx 0
\end{equation}%
corresponding to the lower bandwidth compressed. In the case of $\delta \ll
-2|g|$, the group velocity for the lower photonic band reaches its maximum
\begin{equation}
v_{-}(-\pi /(2\ell ))\approx 2J\ell .
\end{equation}%
And the lower band has a large bandwidth, which accommodates the entire
pulse bandwidth. In Fig.~\ref{fig:2}(a) we have plotted the group velocity
of the lowest band as a function of the level spacing $\omega _{A}$ of the
two-level artificial atom and the coupling strength $g$ at $k=-\pi /(2\ell )$%
. It can be seen that when a large detuning occurs, the group velocity is
minimum at large blue detuning and is maximum at large red detuning. Hence,
for a microwave pulse as a superposition of many $k$-states, its
distribution in the $k$-space can be entirely contained in the energy band
by setting $\delta \ll -2|g|$. Therefore the microwave pulse can be stopped
by adiabatically tuning the detuning from $\delta \ll -2|g|$ to $\delta \gg
2|g|$.

In the near-resonance case $\delta \sim 0$ with strong couplings, the group
velocity in the lower band
\begin{equation*}
v_{-}=\text{Re}\left\{ J\ell \sin (k\ell )\left[ \frac{g}{J\cos (k\ell
)+i\kappa }\right] ^{2}\right\}
\end{equation*}%
is reduced to zero approximately within the range of $k$ satisfying $J\cos
(k\ell )\gg g$. For mode $k$, which satisfies $J\cos (k\ell )\ll -g$, the
corresponding group velocity becomes $v_{-}\approx 2J\ell \sin (k\ell )$.
For mode $k,$ which satisfies $-g\ll J\cos (k\ell )\ll g$, the group
velocity becomes $v_{-}\approx J\ell \sin (k\ell )$, which are illustrated
in Fig.~\ref{fig:2}(b). It tells us that when a microwave pulse inputs into
such a coupled-line-resonator waveguide, some components will be stopped
completely by adjusting the coupling strength $g$, while others still pass
through. So the microwave pulse is distorted. Hence, in the case of
resonance, one cannot obtain the whole information that the wave packet
carries.

\section{\label{sec:four}Coherent output of slow light}

The above boson model describes the linear couplings between two kinds of
boson modes. If one can prepare the state of the charge qubit array with
population inversion through coherent pumping, the dynamic evolution will
drive the coupled-line-resonator mode to output a laser-like light in a
coherent state. Such pumping mechanism with superconducting qubits has been
explored most recently for superconducting flux qubits~\cite{You}.

Now we assume the charge qubits are prepared coherently in excited states.
Then the initial state of the spatially-distributed atomic ensemble is a $k$%
-mode coherent state
\begin{equation}
|\alpha _{k}\rangle =D(\alpha _{k})|G\rangle \equiv D(\alpha
_{k})|g_{1}\rangle \otimes |g_{2}\rangle \otimes \ldots \otimes
|g_{N}\rangle
\end{equation}%
where the displacement operator
\begin{equation}
D(\alpha _{k})=\exp (\alpha _{k}\hat{b}_{k}^{\dag }-\alpha _{k}^{\ast }\hat{b%
}_{k})\text{,}  \label{ctl4-00}
\end{equation}%
describes a superposition of $n$-quasiparticle excitation states $%
|n_{k}\rangle $ of mode-$k$. Here, the one quasiparticle excitation
characterizes the spatially-distributed qubit populations with definite
phases, a quasi-spin wave in the charge qubit array. Let the total system
initially start with a coherent state of atomic ensemble and the vacuum
state of the coupled-line-resonator array, i.e. $|\psi (0)\rangle =|0\rangle
\otimes |\alpha _{k}\rangle $. Here, $|0\rangle $ means no photon contained
in the coupled-line-resonator array. After time $t$, the initial state is
evolved into $|\psi (t)\rangle =U\left( t\right) |\psi (0)\rangle $.

In order to find an explicit expression for $|\psi (t)\rangle $, we formally
rewrite the quantum state at time $t$ as
\begin{equation}
|\psi (t)\rangle =U\left( t\right) D(\alpha _{k})U^{-1}\left( t\right)
U\left( t\right) |G\rangle |0\rangle   \label{ctl4-01a}
\end{equation}%
where $U\left( t\right) =\exp (it\sum_{k}H_{k})$ is the time evolution
operator. \ Because the number of excitations in the total system is
conserved and $|G\rangle |0\rangle $ is the ground state of the system
corresponding to the zero eigenvalue, the quantum state defined by Eq. (\ref%
{ctl4-01a}) becomes
\begin{equation}
|\psi (t)\rangle =U\left( t\right) D(\alpha _{k})U^{-1}\left( t\right)
|G\rangle |0\rangle \text{,}  \label{ctl4-02a}
\end{equation}%
which is completely determined by the time dependent displacement operator
\begin{equation}
D\left( t\right) \equiv U\left( t\right) D(\alpha _{k})U^{-1}\left( t\right)
=\exp \left[ \hat{A}\left( -t\right) \right]   \label{ctl4-03a}
\end{equation}%
where $\hat{A}\left( -t\right) =U\left( t\right) \hat{A}U^{-1}\left(
t\right) $.

With respect to the polariton operators $\hat{P}$ and $\hat{Q}$, the initial
coherent state $|\alpha _{k}\rangle $ of the $k$th mode can be rewritten
based on the displacement operator $D(\alpha _{k})=\exp \left[ \hat{A}(0)%
\right] $, where
\begin{equation}
\hat{A}(0)=\alpha _{k}\hat{P}_{k}^{\dag }\sin \theta _{k}-\alpha _{k}\hat{Q}%
_{k}^{\dag }\cos \theta _{k}-h.c.  \label{ctl4-00a}
\end{equation}%
is an anti-Hermitian operator. Since polaritons are the eigenexcitation of
the total system, their creation operators $\hat{P}_{k}^{\dag }$ and $\hat{Q}%
_{k}^{\dag }$ evolve according to the eigenfrequencies, i.e.,
\begin{eqnarray*}
U\left( t\right) \hat{P}_{k}^{\dag }U^{-1}\left( t\right)  &=&\hat{P}%
_{k}^{\dag }e^{-i\left( \Omega _{Dk}+\varepsilon _{k}\right) t}\text{,} \\
U\left( t\right) \hat{Q}_{k}^{\dag }U^{-1}\left( t\right)  &=&\hat{Q}%
_{k}^{\dag }e^{-i\left( \Omega _{Dk}-\varepsilon _{k}\right) t}\text{.}
\end{eqnarray*}%
Accordingly, the operator $\hat{A}(t)$ can be explicitly obtained as
\begin{eqnarray}
\hat{A}\left( t\right)  &=&\alpha _{k}P_{k}^{\dag }e^{-i\left( \Omega
_{Dk}+\varepsilon _{k}\right) t}\sin \theta _{k} \\
&&+\alpha _{k}^{\ast }Q_{k}e^{i\left( \Omega _{Dk}-\varepsilon _{k}\right)
t}\cos \theta _{k}-h.c.  \notag
\end{eqnarray}%
Transformed back to the original representation with operators $\hat{a}_{k}$
and $\hat{b}_{k}$, the displacement operator $D\left( t\right) $ becomes the
product of two displacement operators, i.e.,
\begin{equation}
D\left( t\right) =D\left[ \alpha _{k}\left( t\right) \right] D\left[ \beta
_{k}\left( t\right) \right] \text{,}  \label{ctl4-03}
\end{equation}%
where the factor
\begin{equation}
D\left[ \alpha _{k}\left( t\right) \right] =\exp \left[ \alpha _{k}\left(
t\right) b_{k}^{\dag }-\alpha _{k}^{\ast }\left( t\right) b_{k}\right]
\label{ctl4-04}
\end{equation}%
acts on the atomic excitation state while
\begin{equation}
D\left[ \beta _{k}\left( t\right) \right] =\exp \left[ \beta _{k}\left(
t\right) a_{k}^{\dag }-\beta _{k}^{\ast }\left( t\right) a_{k}\right]
\label{ctl4-05}
\end{equation}%
acts on the state space of the coupled-line-resonator array.

Therefore, the state $|\psi (t)\rangle $ can be factorized as $|\psi
(t)\rangle =$ $|\beta _{k}(t)\rangle \otimes |\alpha _{k}(t)\rangle $ with
the time-dependent amplitudes
\begin{subequations}
\label{ctl4-04a}
\begin{eqnarray}
\alpha _{k}\left( t\right)  &=&\alpha _{k}e^{-i\Omega _{Dk}t}\left(
e^{i\varepsilon _{k}t}\cos ^{2}\theta _{k}+e^{-i\epsilon _{k}t}\sin
^{2}\theta _{k}\right) \text{,} \\
\beta _{k}\left( t\right)  &=&-ie^{-i\Omega _{Dk}t}\alpha _{k}\sin \left(
\varepsilon _{k}t\right) \sin \left( 2\theta _{k}\right) \text{,}
\end{eqnarray}%
Here the coherent state $|\beta _{k}(t)\rangle $ localizes in the $k$-mode,
and it actually is a spatially multi-mode coherent state; the periodically
modulated amplitudes $\beta _{j}(t)$ mean a quasi-classical wave packet of
photons with distribution $P_{j}=|\beta _{j}(t)|^{2}$. The above argument
means that the initial coherent input of the atomic excitation can result in
a coherent output in the photonic mode, which is characterized by a coherent
state $|\beta _{k}(t)\rangle $. This just displays a laser-like behavior for
the photons output from the coupled line-resonator waveguide.

\section{\label{sec:five}Laser like process with population inversion}

The above intuitive discussion shows an obvious laser behavior, but we need
the population inversion implemented by some coherent pumping. We also need
to consider the threshold condition for the lasing in the coupled
line-resonator waveguide. To this end we study the coherent radiation in the
coupled line-resonator waveguide stimulated by the the artificial atoms with
a collective coherent excitation.

Ignoring the fluctuations due to the couplings to the thermal bath, the
dynamic variable of this system obeys the following equations
\end{subequations}
\begin{subequations}
\label{Laser-01}
\begin{eqnarray}
\partial _{t}\hat{a}_{k} &=&-i(\Omega _{k}-i\eta )\hat{a}_{k}-ig\hat{B}_{k},
\\
\partial _{t}\hat{B}_{k} &=&-i(\omega _{A}-i\gamma )\hat{B}_{k}+i\frac{g}{N}%
\sum_{k^{\prime }}\hat{S}_{k^{\prime }-k}\hat{a}_{k^{\prime }}, \\
\partial _{t}\hat{S}_{0} &=&\Gamma \left( Nd_{0}-\hat{S}_{0}\right)
-i2g\sum_{k}\left( \hat{B}_{k}^{\dag }\hat{a}_{k}-\hat{a}_{k}^{+}\hat{B}%
_{k}\right)
\end{eqnarray}%
where
\end{subequations}
\begin{equation*}
\hat{S}_{k^{\prime }-k}=\sum_{j=0}^{N-1}e^{i\ell \left( k-k^{\prime }\right)
j}\sigma _{j}^{z}
\end{equation*}%
and $\hat{S}_{0}$ is defined by $k^{\prime }=k$. By neglecting the
scattering from $k$ to $k^{\prime }$ due to large $N$, one can write down a
system of laser-like equations~\cite{haken}
\begin{align}
\partial _{t}\hat{a}_{k}& =-i(\Omega _{k}-i\eta )\hat{a}_{k}-ig\hat{b}_{k},
\notag \\
\partial _{t}\hat{b}_{k}& =-i(\omega _{A}-i\gamma )\hat{b}_{k}+ig\hat{n}\hat{%
a}_{k}/N,  \label{Laser} \\
\partial _{t}\hat{n}& =\Gamma \left( Nd_{0}-\hat{n}\right)
-i2g\sum_{k}\left( \hat{b}_{k}^{\dag }\hat{a}_{k}-\hat{a}_{k}^{+}\hat{b}%
_{k}\right) ,  \notag
\end{align}%
where $d_{0}$ is the input rate for equilibrium inversion. The term $\Gamma
\left( Nd_{0}-\hat{n}\right) $ in Eq.~(\ref{Laser}) is phenomenologically
introduced to characterize the role of pumping by some population inversion
for the excitation number $\hat{n}=\sum \hat{b}_{k}^{\dag }\hat{b}_{k}$; $%
\Gamma $ is the relaxation time to equilibrium.

Next, we eliminate the fast time-dependence in Eq.~(\ref{Laser}) by the
substitutions
\begin{equation}
\hat{a}_{k}=\hat{\tilde{a}}_{k}\exp (-i\Omega _{k}t)  \label{ctl4-01}
\end{equation}%
and
\begin{equation}
\hat{b}_{k}=\hat{\tilde{b}}_{k}\exp (-i\Omega _{k}t)\text{.}  \label{ctl4-02}
\end{equation}%
We lock the cavity mode in the coupled-line-resonator in a resonance
frequency $\Omega _{k}\approx \omega _{A}$, that is, $\exp [i(\omega
_{A}-\Omega _{k})]$ varies slowly. When the relaxation time of charge-qubits
is much smaller than the relaxation time of the cavity as well as that of
the population inversion, i.e., $\gamma \gg \Gamma \gg \eta $, we can
adiabatically eliminate $\hat{b}_{k}$ and $\hat{n}$ by setting the
corresponding time derivatives to zero in the obtained equations about $\hat{%
\tilde{a}}_{k}$, $\hat{\tilde{b}}_{k}$ and $\hat{n}$ from Eq.~(\ref{Laser}).
Finally we obtain the equation of  motion for $\hat{\tilde{a}}_{k}$
\begin{equation}
\partial _{t}\hat{\tilde{a}}_{k}=(d_{0}g^{2}L-\eta )\hat{\tilde{a}}_{k}-%
\frac{4\gamma d_{0}g^{4}L\left\vert L\right\vert ^{2}}{N\Gamma }\hat{\tilde{a%
}}_{k}^{\dag }\hat{\tilde{a}}_{k}\hat{\tilde{a}}_{k},  \label{laser1}
\end{equation}%
where the Lorentz distribution
\begin{equation}
L=\frac{1}{\gamma +i(\omega _{A}-\Omega _{k})}
\end{equation}%
shapes the spectra of coherent output. It is exactly the laser equation \cite%
{haken} defining a threshold $d_{0}=\eta /(g^{2}L)$ for the laser-like
output in the coupled-line-resonator waveguide by the coherent pumping with
an input rate $d_{0}$.

\begin{figure}[tbp]
\includegraphics[width=5.5 cm]{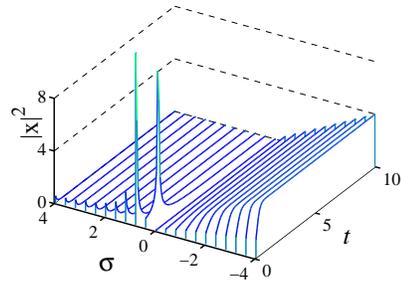}
\caption{(Color online) The probability amplitude $|x|^{2}$ of
output photon as a function of $\protect\sigma $ and time. When
$\protect\sigma <0$, there is a laser-like output depicted by the
stable solution for $t\rightarrow \infty$. $t$ is in units of
seconds ``s'' and $\protect\sigma$ is in units of s$^{-1}$.}
\label{fig:laser}
\end{figure}

This dynamical lasing behavior can be described by the non-linear equation
\begin{equation}
\overset{\cdot }{x}=-\sigma x-\lambda x^{3}  \label{ordern}
\end{equation}%
about the order parameters $x=\langle \alpha |\hat{\tilde{a}}_{k}|\alpha
\rangle $. Here, $|\alpha \rangle $ is a coherent state with real number $%
\alpha $; the coefficient of the linear term of Eq. (\ref{ordern})
\begin{equation}
\sigma =\eta -d_{0}g^{2}L
\end{equation}%
describes the amplification effect of the light field when $\sigma <0$; the
coefficient
\begin{equation}
\lambda =\frac{4\gamma d_{0}g^{4}L\left\vert L\right\vert ^{2}}{N\Gamma }
\end{equation}%
represent the nonlinearity of the effective theory obtained by averaging the
atomic excitations. The nonlinear coefficient competes with the parameter $%
\sigma $ to realize a lasing \textquotedblleft phase
transition\textquotedblright . As illustrated in Fig.~\ref{fig:laser}, the
solutions of the non-linear equation obviously possess a critical behavior
near the threshold $\sigma =0$ . The solution of Eq.~(\ref{ordern}) gives
\begin{equation}
\left\vert x\right\vert ^{2}=[c\exp (2\sigma t)-\lambda /\sigma ]^{-1}
\end{equation}%
for $\sigma \neq 0$; and
\begin{equation}
|x|^{2}=(2\lambda t+c)^{-1}
\end{equation}%
for $\sigma =0$, where $c$ is a constant determined by the initial state.
The stable solutions that
\begin{equation}
\left\vert x\right\vert ^{2}=-\sigma /\lambda
\end{equation}%
for $\sigma <0$ (above the threshold) and $\left\vert x\right\vert ^{2}=0$
for $\sigma >0$ (below the threshold) mean a laser-like output of the
coupled-line-resonators. Since the expectation value of $\hat{\tilde{a}}_{k}$
does not vanish, we can understand the above laser-like effect as a
pumping-induced symmetry-breaking.

Together with the intuitive argument in the last section, the analysis in
this section definitely shows the existence of a threshold for laser-like
behavior in the present artificial system. However, to reduce the threshold
by overcoming decoherence, including dissipation and dephasing, is still a
challenge to implement a laser of slow light in practical hybrid systems.

\section{\label{sec:six}Concluding remarks}

In conclusion, we have conceptually proposed an electromagnetically
controllable quantum device based on superconducting circuit QED for the
coherent manipulation of photons. It can realize a laser-like slow microwave
output from a coupled-line-resonator waveguide by controlling each cavity
connected to a charge qubit. Our studies are motivated by the all-optical
experiment for stopping light in a coupled resonator waveguide, which is
coupled to other cavities, but here we show that some results for the
coupled resonator waveguide can be obtained by replacing the coupling
cavities with a more practical system, a spatially-distributed array of
charge qubits. Finally, we would like to mention that, without the
limitation of the superconducting system, the present analysis represents a
universal setup for the coherent manipulation for light or microwave
propagations in some all-optical or electromagnetic-optical system.

This work is supported by the NSFC with Grants No. 90203018,
10474104, 60433050, 10547101, 10604002 and 10704023, NFRPC with
Grant No. 2001CB309310, No. 2005CB724508, No. 2006CB921205, and K.
C. Wong Education Foundation. We acknowledge the useful discussions
with W. Li and Jing Lu.


\begin{thebibliography}{99}
\bibitem{Fan1} Q. Xu, S. Sandhu, M. L. Povinelli, J. Shakya, S. Fan, and M.
Lipson, Phys. Rev. Lett. \textbf{96}, 123901 (2006).

\bibitem{Fan2} M. F. Yanik, W. Suh, Z. Wang, and S. Fan, Phys. Rev. Lett.
\textbf{93}, 233903 (2004).

\bibitem{Fan3} M. F. Yanik and S. Fan, Phys. Rev. Lett. \textbf{92}, 083901
(2004).

\bibitem{Fan4} R. W. Boyd and D. J. Gauthier, Nature \textbf{441}, 701
(2006).

\bibitem{EIT97} S. E. Harris, Phys. Today \textbf{50}, No. 7, 36 (1997).

\bibitem{EIT01} L.V. Hau, S.E. Harris, Z. Dutton and C.H. Behroozi, Nature
\textbf{397} 594 (1999).

\bibitem{sunprl91} C. P. Sun, Y. Li and X. F. Liu, Phys. Rev. Lett. \textbf{%
91}, 147903 (2003).

\bibitem{YouPT58} J. Q. You and F. Nori, Phys. Today \textbf{58} (11), 42
(2005).

\bibitem{YouPRB68} J. Q. You and F. Nori, Phys. Rev. B \textbf{68}, 064509
(2003).

\bibitem{YouJPRB68} J. Q. You, J. S. Tsai and F. Nori, Phys. Rev. B \textbf{%
68}, 024510 (2003).

\bibitem{NTTPRL96} J. Johansson, S. Saito, T. Meno, H. Nakano, M. Ueda, K.
Semba, and H. Takayanagi, Phys. Rev. Lett. \textbf{96}, 127006 (2006).

\bibitem{mn431} I. Chiorescu, P. Bertet, K. Semba, Y. Nakamura, C. J. P. M.
Harmans, and J. E. Mooij, Nature \textbf{431}, 159 (2004)

\bibitem{yale1} A. Wallraff, D. I. Schuster, A. Blais, L. Frunzio, R. S.
Huang, J. Majer, S. Kumar, S. M. Girvin, R. J. Schoelkopf, Nature \textbf{431%
}, 162 (2004);

\bibitem{yale2} A. Blais, R.-S. Huang, A. Wallraff, S. M. Girvin and R. J.
Schoelkopf, Phys. Rev. A \textbf{69}, 062320 (2004).

\bibitem{liuPRA71} Y. X. Liu, L. F. Wei, and F. Nori, Phys. Rev. A \textbf{71%
}, 063820 (2005)

\bibitem{liuEL67} Y. X. Liu, L. F. Wei, and F. Nori, Europhys. Lett \textbf{%
67}, 941 (2004)

\bibitem{nec2q1} Y. A. Pashkin, T. Yamamoto, O. Astafiev, Y. Nakamura, D.V.
Averin and J.S. Tsai, Nature (London), \textbf{421}, 823 (2003);

\bibitem{nec2q2} T.Yamamoto, Yu. A. Pashkin, O. Astafiev, Y. Nakamura, and
J. S. Tsai, Nature (London), \textbf{425}, 941 (2003).

\bibitem{4q} M. Grajcar, A. Izmalkov, S. H. W. van der Ploeg, S. Linzen T. Plecenik, T.
Wagner, U. H\"{u}bner, E. Il'ichev, H.-G. Meyer, A. Y. Smirnov, P.
J. Love, A. Maassen van den Brink, M. H. S. Amin, S. Uchaikin, and
A. M. Zagoskin, Phys. Rev. Lett. \textbf{96}, 047006 (2006).

\bibitem{nec99} Y. Nakamura, Yu.A. Pashkin and J.S. Tsai, Nature \textbf{398}%
, 786 (1999).

\bibitem{GWe0508} G. Wendin and V. S. Shumeiko, e-print
arXiv:cond-mat/0508729.

\bibitem{Bqp06159} D. G. Angelakis, M. F. Santos, and S. Bose, Phys.
Rev. A \textbf{76}, 031805(R) (2007)

\bibitem{Pqp06097} M. J. Hartmann, F. G. S. L. Brandao and M. B. Plenio,
Nat. Phys. \textbf{2}, 849 (2006);

\bibitem{Dtree} D. Greentree, C. Tahan, J. H. Cole, and L. C. L. Hollenberg,
Nat. Phys. \textbf{2}, 856 (2006).

\bibitem{Wang} Y. D. Wang, Fei Xue, and C. P. Sun, e-print arXiv:
quant-ph/0603014.

\bibitem{Quan} H. T. Quan, Z. Song, X. F. Liu, P. Zanardi and C. P. Sun,
Phys. Rev. Lett. \textbf{96}, 140604 (2006).

\bibitem{You} J. Q. You, Y.-X. Liu, C. P. Sun, and F. Nori, Phys. Rev. B
\textbf{75}, 104516 (2007).

\bibitem{haken} H. Haken, Rev. Mod. Phys. \textbf{47}, 67 (1975).
\end{thebibliography}
\end{document}